\documentclass[twocolumn,aps,floats,showpacs,superscriptaddress]{revtex4}
\usepackage{amsmath}
\usepackage{bbm}
\usepackage{graphicx}

\newcommand{\tr}[1]{{{\textrm{#1}}}}
\newcommand{\nn}{\nonumber}

\begin{document}

\bibliographystyle{apsrev}

\title{Interplay between exchange interactions and charging effects in metallic grains.}
\author{Pablo San-Jose}
\email{pablo.sanjose@icmm.csic.es}
 \affiliation {Instituto de Ciencia de Materiales de Madrid. CSIC. Cantoblanco. 28049 Madrid. Spain}
 \author{Carlos P. Herrero}%
\email{ch@icmm.csic.es}
 \affiliation {Instituto de Ciencia de
Materiales de Madrid. CSIC.
Cantoblanco. 28049 Madrid. Spain} %
\author{Francisco Guinea}%
\email{paco.guinea@icmm.csic.es} \affiliation {Instituto de
Ciencia de Materiales de Madrid. CSIC. Cantoblanco. 28049 Madrid.
Spain}
\author{Daniel P. Arovas}%
\email{darovas@ucsed.edu}%
\affiliation{Department of Physics, University of California at
San Diego, La Jolla California 92093, USA}

\begin{abstract}
We study the effect of the exchange interactions in small superconducting grains near the
Coulomb blockade regime. We extend the standard description of the grain in terms of a
single collective variable, the charge and its conjugated phase, to include the spin
degree of freedom. The suppression of spin fluctuations enhances the tendency towards
Coulomb blockade. The effective charging energy and conductance are calculated
numerically in the regime of large grain-lead coupling.
\end{abstract}
\maketitle

{\it Introduction.} Coulomb blockade in metallic grains is a well studied phenomenon
\cite{AL91,SET}.  The transport through a grain in the Coulomb blockade regime can be
studied using rate equations when the coupling to the leads is weak \cite{AL91,SZ90}. The
renormalization of the charging energy when the coupling to the leads is large is also
well understood \cite{SZ90,GS86,PZ91,HSZ99}. The strongly coupled regime is best studied
by introducing a single collective degree of freedom, the phase, conjugated to the total
number of electrons in the grain $Q/e$ \cite{AES82,H83,BMS83}.  The use of this variable
is justified when the separation between the electronic levels within the grain can be
neglected, or, alternatively, when the conductance of the grain is large. In this limit,
the interaction effects within the grain can be described by a simple hamiltonian
\cite{KAA00}, expressed in terms of the total charge ($E_C (\hat{Q}-Q_0)^2$), the total
spin ($J_S{\vec S}^2$) and the individual electronic degrees of freedom.  {In the
presence of attractive interactions in the grain a pairing term
($\lambda_{BCS}\hat{T}^+\hat{T}$) should also be included \cite{Aleiner02} that will
drive the system towards superconducting state with energy gap $\Delta$. In such case
both the pairing $\lambda_{BCS}$ and the exchange $J_S$ will grow under renormalization
group (integration of energies down from Thouless energy) to a scale much larger than the
bare one, which is initially of the order of the level spacing. Moreover due to the
attractive character of interaction the spin susceptibility due to exchange will be
positive, so that spin fluctuations get suppressed much in the same way as charge
fluctuations do due to the charging energy. The essential distinction between the two
will stem from the topological differences between spin group SU(2) and charge group U(1)
in which their conjugate phases exist.}

In the following, we will generalize the usual description of
{a small grain in the Coulomb blockade regime in terms of
phase dynamics \cite{AES82}}. We include also the dynamics of the
total spin of the grain {in SU(2)}, on the same footing as
the total charge. {We assume that the grain has a
negligible level spacing and a finite positive renormalized
susceptibility according to the arguments above.}

 {The coupling between the grain and an external (normal)
electrode leads to single electron tunnelling. We describe it as
usually done, in terms of a long range interaction, in time,
between the phase and spin variables of the
grain \cite{EckernPRB84}. In most of the calculations, we neglect
the long time cutoff in the kernel describing these interactions
imposed by the superconducting gap. This approximation is
reasonable when the bare charging energy is larger than the gap,
and its renormalization arises from virtual tunnelling processes
of high energy. The effects arising from the suppression of the
subgap conductance due to the superconducting state can be
incorporated into our calculations in a straightforward way.
Finally, our method can be used to study superconducting junctions
with subgap leakage currents \cite{Ietal89}, although we will not
consider this case in detail.}

{The effects of a constant exchange term on the transport properties of a quantum
dot has already been studied in the limit where the coupling to the leads is weak, using
rate equations \cite{Aetal02,AR03,UB03}. The present formalism goes one step beyond this
by summing all processes up to cotunneling level}.However we do not consider here
the changes in the grain susceptibility induced by the spin-orbit coupling which has been
considered by other authors \cite{MGL00}.

{\it The model.} {As mentioned above, we will focus on the
case in which the superconducting gap is smaller or of similar
order as the renormalized charging energy, $E_C^*$, and analyze
how this renormalization of the bare charging energy, $E_C$,
depends on the exchange J. These constraints are satisfied, for
instance, by Al superconducting grains with radii below $100$ nm.}

The hamiltonian that we study is: ${\cal H} ={\cal H}_{\rm grain}
+ {\cal H}_{\rm lead}+ {\cal H}_{\rm hop}$, where
\begin{eqnarray}
{\cal H}_{\rm grain}&= &\sum_{i,s} \epsilon^{\vphantom\dag}_i d^\dag_{i,s} d_{i,s} + E_C
{\hat N}^2 + J_S {\vec S}^2 \nonumber \\
{\cal H}_{\rm lead} &= &\sum_{k,s} \epsilon^{\vphantom\dag}_k c^\dag_{k,s}
c^{\vphantom\dag}_{k,s}
\nonumber
\\ {\cal H}_{\rm hop} &= &-t \sum_{i,k,s} c^\dag_{i,s} d^{\vphantom\dag}_{k,s} + h. c.
\label{hamil}
\end{eqnarray}
and $\hat{N}$ and $\vec{S}$ are the total number of electrons and
the total spin of the grain, $\hat{N} = \sum_{i,s} d^\dag_{i,s}
d^{\vphantom\dag}_{i,s}$ and $\vec{S} = \frac{1}{2}\sum_{i,s,s'}
d^\dag_{i,s}{\vec\sigma}^{\vphantom\dag}_{s,s'}
d^{\vphantom\dag}_{i,s'}$, where $\vec{\sigma}$ denotes the Pauli
matrices. The grain-dot conductance, in dimensionless units, can
be approximated by $\alpha \approx t^2 \rho_{\rm grain} (
\epsilon_{\rm F} ) \rho_{\rm lead}( \epsilon_{\rm F})$, where
$\rho_{\rm grain/lead} ( \epsilon_{\rm F})$ is the density of
states at the Fermi level of the grain and the lead. {As
mentioned above, we neglect the energy dependence of the density
of states of the grain.} The only interactions included in
eq.(\ref{hamil}) are through the total spin and charge of the
grain.

{\it Path integral formulation.} We can integrate out the
fermionic degrees of freedom and obtain a description in terms of
collective variables only by using the path integral formalism.
The action is $S=S_{\rm grain}^0+S_{\rm lead}+S_{\rm hop}+S_{\rm int}$, where
\begin{eqnarray}
S_{\rm grain}^0&=&\int_0^\beta\!\! d \tau \sum_{i,s}  {\bar d}_{i,s}\,(\partial_\tau+\epsilon_i-
\mu_{\rm grain})\,d_{i,s}\nn\\
S_{\rm lead}&=&\int_0^\beta \!\! d \tau \sum_{k,s} {\bar c}_{k,s}\,(\partial_\tau+
\epsilon_k-\mu_{\rm lead})\,c_{k,s}\nn\\
S_{\rm hop}&=&-t \int_0^\beta d \tau \sum_{i,k,s} {\bar d}_{i,s}\,c_{k,s}+{\rm H.c.}\nn\\
S_{\rm int}&=&E_C(\hat{N}-N_\tr{ext})^2+J_S
(\vec{S}-\vec{S}_\tr{ext})^2
\end{eqnarray}
We have included an offset electron number $N_\tr{ext}=V_{\rm gate}/eC_g$ induced by a
gate voltage $V_{\rm gate}$ and an offset spin $\vec{S}_\tr{ext}=\vec{H}_\tr{ext}/2J_S$
induced by an external magnetic field $\vec{H}_\tr{ext}$, that couples to the total spin.

We can now decouple the quartic interaction term $S_\tr{int}$ by
means of a Hubbard-Stratonovich transformation
$e^{-Ec(N-N_\tr{ext})^2}\propto\int\mathcal{D}V e^{-V^2/4E_\tr{C}-i V
(N-N_\tr{ext})}$, and similarly for $\vec{S}$, which introduces a
new scalar field $V$ for the total charge and a \emph{vector}
field $\vec{H}$ for the total spin.  We then have ${\cal S}_{\rm int}=S_0+S_1$, with
\begin{eqnarray}
S_0&=&\int_0^\beta\!\! d\tau\left(\frac{V^2}{4E_C}+\frac{\vec{H}^2}{4J_S}-
i VN_\tr{ext}-i\vec{H}\cdot\vec{S}_\tr{ext}\right)\nn \label{SHV}\\
S_1&=&i\int_0^\beta\!\! d\tau \left(V \hat{N} +
\vec{H}\cdot\vec{S}\right)\ .
\end{eqnarray}
We now perform a time dependent canonical transformation (a phase and spin rotation) on
the electronic wavefunctions, in order to cancel the term $S_1$ in
eq.(\ref{SHV}).  This $U(1)\times SU(2)$ transformation can be written as:
\begin{eqnarray}
d_{ks}(\tau)&\rightarrow &U_{ss'}(\tau)\,d_{ks'}(\tau) \nn \\
U(\tau) &=&e^{i\phi(\tau)}\,e^{\frac{i}{2}\xi(\tau){\hat n}(\tau)\cdot\vec{\sigma}}
\label{transformation}
\end{eqnarray}
The transformation is parametrized by the angles $\phi ( \tau )$
and $\xi ( \tau )$, and by the three dimensional unitary vector
$\hat{n} ( \tau )$.  The requirement that $S_1$ in (\ref{SHV}) is
cancelled implies:
\begin{equation}
(\partial_\tau U)\,U^\dag=iV+\frac{i}{2}\vec{H}\cdot\vec{\sigma}
\end{equation}
so that $V={\dot\phi}$ and
\begin{equation}
\vec{H}=\dot{\xi}\,{\hat n}+\sin\xi\,\dot{\hat n}+
(1-\cos\xi)\,\dot{\hat n}\times{\hat n} \label{VH}
\end{equation}

These identities provide an alternative and convenient
parametrization of the auxiliary fields $V$ and $\vec{H}$,
represented now by the $\phi,\xi,{\hat n}$ fields, which will be
used in the following. Note that the $U(1)$ gauge transformation
needed to replace $V$ by the phase $\phi$ leads to the standard
description of charging effects in terms of phase fluctuations. It
is interesting to note that eq.(\ref{VH}) implies that $\vec{H}$
is proportional to the angular momentum of a \emph{sphere},
considered as a rigid body \cite{LL77}. The periodicity in
imaginary time of the arguments in the action implies that
$U(0)=U(\beta)$. This constraint implies the usual quantization of
the charge in the grain, and also of the spin (see below), due to
the discreteness of transport events.

A more compact notation for the transformation in
eq.(\ref{transformation}) can be given in terms of the following,
$\tau$ dependent, two- and four-dimensional unit vectors:
\begin{eqnarray}
{\hat u}_\tau&=&(\sin\phi_\tau,\cos\phi_\tau)\nn\\
{\hat v}_\tau&=&\left( {\hat n}_\tau
\sin\frac{1}{2}\xi_\tau,\cos\frac{1}{2}{\xi_\tau}\right)
\end{eqnarray}
We can use these vectors to write $S_0$ as:
\begin{eqnarray}
S_0=\int_0^\beta d\tau\left\{\frac{(\partial_\tau {\hat u})^2}{4 E_C}
+\frac{( \partial_\tau {\hat v})^2}{4 J_S}\right\}
\label{S_0}
\end{eqnarray}
where external gates and fields have been taken as zero for the
moment. The transformation in eq.(\ref{transformation}) modifies
also the lead-grain coupling, $S_{\rm hop}$:
\begin{equation}
S_{\rm hop}=-t\int_0^\beta\!\!d\tau \sum_{i,k,s,s'}{\bar c}_{i,s}\,U_{s,s'}\,
d_{k,s'}+\textrm{H.c.}
\end{equation}

A final step to obtain the effective action for the rotor fields
is to integrate out the fermionic fields to order $t^2$, using
$\langle e^{-S_\tr{hop}}\rangle_0= e^{-\frac{1}{2}\langle
S_\tr{hop}^2\rangle_0+\mathcal{O}(t^4)}$. One obtains the
following dissipation term:
\begin{equation}
S_{\rm diss}=-\frac{\alpha}{4}\int_0^\beta\!\!\!d\tau\int_0^\beta\!\!\!d\tau'\,
K(\tau-\tau')\,\tr{Tr}\Big[U^\dag_\tau U^{\vphantom\dag}_{\tau'}
+U^\dag_{\tau'}U^{\vphantom\dag}_{\tau}\Big]\nn
\end{equation}
where $K(\tau)=-[G_{\rm lead}(\tau)G_{\rm grain}(-\tau)]/[\rho_{\rm lead} ( \epsilon_{\rm
F} )\rho_{\rm grain} ( \epsilon_{\rm F}) ]=(\pi T)^2/\sin^2(\pi T \tau)$, $G_{\rm lead}$
and $G_{\rm grain}$ being the lead and grain unperturbed Green's functions in imaginary
time.  {Recall here the that the finite range that the superconducting gap could
bring in is assumed larger than the decay time of the phase correlators which is of order
$E_C^*$, so that the gapless $K(\tau-\tau')$ of the normal state yields equivalent
results}. $S_{\rm diss}$ may be finally recast as
\begin{equation}
S_{\rm diss}=\alpha\int_0^\beta\!\!\!\!\!d\tau\!\!\int_0^\beta\!\!\!\!\!d\tau'\,
 K(\tau-\tau')\Big[1-({\hat u}_\tau\cdot{\hat u}_{\tau'})\,
({\hat v}_\tau\cdot{\hat v}_{\tau'})\Big] \label{Sdis}
\end{equation}

This term is sufficient to account for second order tunnelling
processes, and in particular it can describe cotunnelling
features. The derivation is valid when the conductance
between the grain and the electrode {\it per channel\/} is small,
and it can be used even if the total value is large.

The method leading to eq.(\ref{Sdis}) can be easily generalized to
the case when the density of states in the grains or in the leads
is spin dependent. The effective action will contain terms involving
$\sin ( \xi_\tau + \xi_{\tau'} )$ which break the symmetry
between the four components of the vector ${\hat{v}}_\tau$. These
terms are analogous to the Josephson term which arises in the
charge dynamics when the leads, or the grain are superconductors
\cite{AES82}.

The final action is
$S_{\rm eff} =S_0+S_{\rm diss}$,written in terms of the dynamical variables
${\hat u}_\tau$ and ${\hat v}_\tau$ only. In the limit $J_S = 0$, the field
${\hat{v}}_\tau$ can be taken as a constant, and the model reduces
to the standard phase only model.

{\it Results.} It is instructive to analyze first the decoupled
grain, described by $S_0$ in eq.(\ref{S_0}), to see where this
spherical rotor description of the total spin comes from. As
mentioned above, $S_0$ contains the usual phase term, which leads
to the quantization of the charge, and a contribution which is
equivalent to that of a rigid rotor, and which leads to the
conservation of spin. The eigenvalues associated to $S_0$ can be
written as $E_{N,S,S_z,K} = E_C N^2 + J_S S ( S + 1 )$,
where $N = 0,1,2 \cdots$, $S = 0,1,2 \cdots$, $-S \le S_z \le S$
and $- S \le K \le S$. The degeneracy of a given state is $(2 S +
1 )^2$ \cite{LL77}. This degeneracy can be understood by noting
that, in the limit studied here, the level spacing within the
grain can be neglected. The grain energy is solely determined by
the total charge and the total spin. Let us assume that, in the
neutral dot, there are $N_0$ spin 1/2 electrons which contribute
to the total spin. The number of states of total spin $S$ (each
with degeneracy $2S+1$) is:
\begin{equation}
C^{N_0}_{S}=\left(\begin{array}{c}N_0\\\frac{N_0}{2}-S\end{array}\right)-
\left(\begin{array}{c}N_0\\\frac{N_0}{2}-S-1\end{array}\right)
\end{equation}
In the limit of many electrons $N_0\rightarrow\infty$,
one obtains $\lim_{N_0/S\rightarrow\infty} C^{N_0}_{S}=(2S+1) C_{N_0}$, where
$C_{N_0}=\frac{2^{N_0+3/2}}{\sqrt{\pi}N_0^{3/2}}$ is a constant independent of $S$. This
means that the total degeneracy of a state composed of many $1/2$ spins and given value
of the total spin momentum $\langle \hat{S}^2\rangle=S(S+1)$ is $C_{N_0} (2S+1)^2$, just
as $C_{N_0}$ rigid rotors with total angular momentum $S$. The existence of this
degeneracy leads to a prefactor in the free energy which is independent of the angular
momentum. This multiplicity, like similar degeneracies in the case of ordinary Coulomb
blockade, does not affect the effects associated to the spin gap discussed in this paper.

The following calculations including the full action $S_{\rm eff}$
have been done by averaging over all paths in the unit circle
parameterized by ${\hat{u}}$ and the four-dimensional sphere which
defines ${\hat{v}}$, using an extension of the Monte Carlo code
developed earlier for related problems \cite{HSZ99,Betal00}. The
effective charging energy is calculated by summing over winding
numbers of the phase. The conductance between the grain and
the electrode has been approximated by the expression $G ( \beta
/ 2 )$ \cite{ZS91,Betal00}, valid at low temperatures, where $G$
is the correlation function, in imaginary time, of the variable
${\hat{v}}_\tau {\hat{u}}_\tau$. The latter combination describes
the transfer of a full electron to the grain. We calculate,
separately, the correlations $G_u=\langle {\hat u}_\tau\cdot {\hat
u}_{\tau'} \rangle$ and $G_v=\langle {\hat v}_\tau\cdot {\hat
v}_{\tau'} \rangle$ which correspond to charge only and spin only
currents.

\begin{figure}
\includegraphics[height=5cm]{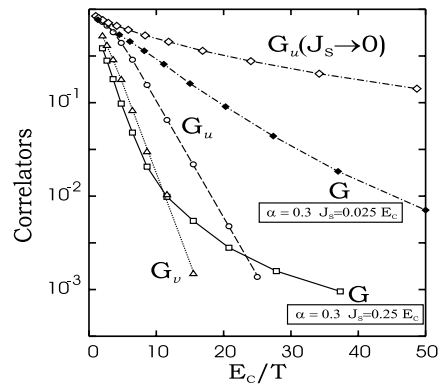}
\caption{Charge-charge, spin-spin and electron-electron current correlations
(see text) versus inverse
temperature for $\alpha=0.3$ and different values of $J_S$.} \label{gbeta}
\end{figure}
\begin{figure}
\includegraphics[height=5cm]{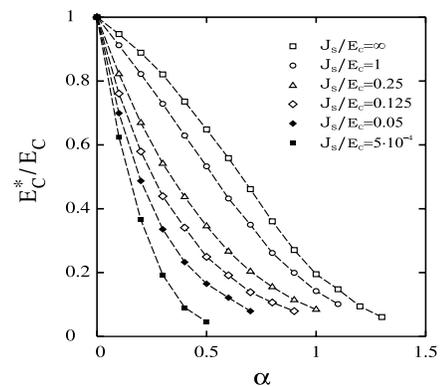}
\caption{Renormalized charging energy of the grain in the presence
of a finite spin gap $J_S$, versus the dimensionless grain-lead
coupling $\alpha$. Note that the decay becomes less pronounced for
growing spin gap.} \label{Ecalpha}
\end{figure}
\begin{figure}
\includegraphics[height=5cm]{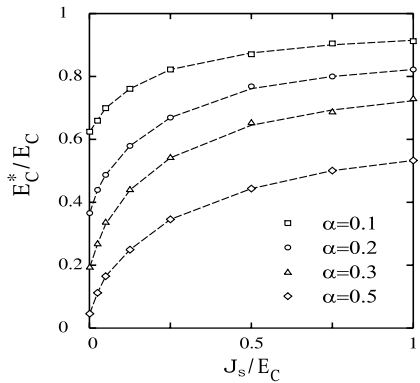}
\caption{Renormalized charging energy of the grain in the presence
of a finite spin gap $J_S$, versus the value of the spin gap
$J_S$. Note the saturation for large $J_S$.} \label{EcJs}
\end{figure}
The current correlation functions are shown in Fig.[\ref{gbeta}].
It is interesting to note that both $\langle {\hat u}_\tau\cdot
{\hat u}_{\tau'} \rangle$ and $\langle {\hat v}_\tau\cdot {\hat
v}_{\tau'} \rangle$ decay exponentially, while the composite
correlation $\langle {\hat u}_\tau \cdot {\hat u}_{\tau'}\, {\hat
v}_\tau\cdot {\hat v}_{\tau'} \rangle$ decays as $( \tau - \tau'
)^{-2}$, as required by Griffith's inequality \cite{G67}.   The
differences between the phase-phase, ``rotation-rotation'' and
current-current correlations is reminiscent of the behavior of a
Luttinger liquid. It implies that the electron current cannot be
factorized into its spin and charge components. The exponential
decay of the correlations associated with the collective charge
and spin degrees of freedom can be understood as the effect of a
charge and spin gap in the grain. It can be obtained by making a
mean field decoupling of the variables, in a similar way to the
calculation for charging effects in coupled grains \cite{Aetal03}.
The $( \tau - \tau')^{-2}$ decay of the current-current
correlation describes the cotunnelling processes at low
temperatures.

The effective charging energies, as functions of $\alpha$ and $J_S$, are shown in
Figs.[\ref{Ecalpha}] and [\ref{EcJs}]. The effect of a finite $J_S$ on the renormalized
charging energy is significant, even for small values of $J_S$. We can estimate
analytically this effect, by assumming that when $J_S \rightarrow 0$ the fluctuations in
the variable ${\hat{v}}_\tau$ are small. The effect of these fluctuations on the variable
${\hat{u}}_\tau$ can be approximated by replacing $\alpha$ in eq.(\ref{Sdis}) by
$\alpha\, \langle | {\hat v} |^2 \rangle$. Assuming that the fluctuations of ${\hat
v}_\tau$ are harmonic, we find:
\begin{equation}
\langle | {\hat v} |^2 \rangle \approx 1 - \int_{E_C}^{\Lambda} \frac{d \omega} {\omega^2
/ 2 J_S} \approx 1 - \frac{J_S}{E_C}
\end{equation}
where $\Lambda$ is a high energy cutoff, comparable to the electronic bandwidth. Then,
using the well known expression for the renormalized charging energy for large values of
$\alpha$ \cite{GS86,SZ90}: $E_C^* \approx E_C^{\vphantom{*}}\,\exp\left\{- 2 \pi^2 \alpha
\left( 1 - \frac{J_S}{E_C} \right)\right\} $. This enhancement of the effective charging
energy by a spin gap is another manifestation of the non-separability of charge and spin.

{\it Conclusions} -- We have analyzed the influence of the exchange term in a small
superconducting grain on the charging effects in the regime where the superconducting gap
is smaller or comparable to the charging energy. The suppression of the spin
susceptibility reduces large fluctuations in the spin of the grain, and enhances the
tendency towards Coulomb blockade. Our analysis integrates out the electronic degrees of
freedom in the grain and in the external leads, and provides a simple description in
terms of the charge and spin degrees of freedom of the grain only.

The effects of the exchange term have been analyzed for closed quantum dots, which are
almost decoupled from the leads \cite{Aetal02,AR03,UB03}. Our scheme provides a
generalization which is {non-perturbative} in the coupling strength {in the
sense that one can recover exponential effects in the coupling, such as the
renormalization of the charging energy, which cannot be derived from the addition of
sequential processes}.

A statistical approximation to the electron-electron
interactions in a small dot predicts that the bare exchange term
is negative and of the order of the separation between electronic
levels \cite{KAA00,AleinerPR02}. Our analysis, on the other hand,
is valid only when the exchange term is positive and larger than
the level spacing. This regime corresponds to systems with an
attractive electron-electron interaction near a superconducting
transition, when the exchange $J$ is significantly
enhanced \cite{MSD72}. Spin fluctuations in a superconducting grain
at low temperatures can therefore have a strong influence on
charge fluctuations, restoring the system to a Coulomb blockade
regime even when the coupling to the leads is strong.

{\it Acknowledgements} -- Two of us (P. S. J. and F. G.) are thankful
to MCyT (Spain) for financial support through grant
MAT2002-0495-C02-01.
\bibliography{Biblio}
\end{document}